\documentstyle[12pt]{article}
\begin{document}
\title{QUANTUM, CHAOS AND THE UNIVERSE}
\author{B.G. Sidharth$^*$\\
Centre for Applicable Mathematics \& Computer Sciences\\
B.M. Birla Science Centre, Adarsh Nagar, Hyderabad - 500063 (India)}
\date{}
\maketitle
\footnotetext{\noindent E-mail:birlasc@hd1.vsnl.net.in}
\begin{abstract}
In this paper we suggest a formulation that would bear out the spirit of
Prigogine's "Order Out of Chaos" and Wheeler's "Law Without Law". In it a
typical elementary particle length, namely the pion Compton wavelength arises
from the random motion of the $N$ particles in the universe of dimension
$R$. It is then argued in the light of recent work that this is the origin
of the laws of physics and leads to a cosmology consistent with observation.
\end{abstract}
\section{Introduction}
Although the universe is apparently governed by iron clad laws, it would be
more natural to expect that the underpinning for these laws would be
chaos itself, that is Prigogine's, "Order out of chaos" or
in the words of Wheeler\cite{r1}, we seek
ultimately a "law without law." As he put it, (loc.cit), "All of physics
in my view, will be seen someday to follow the pattern of thermodynamics
and statistical mechanics, of regularity based on chaos, of "law without
law". Specifically, I believe that everything is built higgledy-piggledy
on the unpredictable outcomes of billions upon billions of elementary
quantum phenomena, and that the laws and initial conditions of physics
arise out of this chaos by the action of a regulating principle, the
discovery and proper formulation of which is the number one task...."
A step in this direction has been the considerable amount of work of
Nelson\cite{r2}, De Pena\cite{r3,r4}, Lehr\cite{r5}, Gaveau\cite{r6},
Landau\cite{r7} and others, who have tried to
derive the Schrodinger equation, the Klein-Gordon equation and even the Dirac
equation from stochastic considerations, and in general develop an underpinning
of stochastic mechanics and stochastic electrodynamics. The literature is vast and
some of the references given cite an extensive bibliography. However, all these
derivations contain certain assumptions whose meaning has been unclear.\\
In the spirit of the above considerations, we propose below that purely
stochastic processes lead to minimum space-time intervals of the order of
the Compton wavelength and time, and it is this circumstance that underlies
quantum phenomena and cosmology, and, in the thermodynamic limit in which
$N$, the number of particles in the universe $\to \infty$ classical
phenomena as well. In the process, we will obtain a rationale for the adhoc
assumptions referred to above.
\section{The Emergence of Quantuzed Space-Time and Physics}
Our starting point is the well known fact that in a random walk, the average
distance $l$ covered at a stretch is given by\cite{r8}
\begin{equation}
l = R/\sqrt{N}\label{e1}
\end{equation}
where $R$ is the dimension of the system and $N$ is the total number of
steps. We get the same relation in Wheeler's famous travelling salesman problem
and  similar problems\cite{r9,r10} The interesting fact that equation (\ref{e1})
is true in the universe itself with $R$ the radius of the universe $\sim 10^{28}cm,N$
the number of the particles in the universe $\sim 10^{80}$ and $l$ the Compton
wavelength of the typical elementary particle, the pion $\sim 10^{-13}cm$ has been
noticed\cite{r11,r12}.\\
We mention in passing that equation (\ref{e1}) which has been considered by
some to be accidental, may not be so at all and has been shown to arise quite naturally in a
cosmological scheme based on fluctuations. As this has been discussed by the
author in detail at other places, for example\cite{r13,r14} the details will
not be given, but we will merely touch upon the main results shortly.\\
We would like to stress that the Compton wavelength given by (\ref{e1}) is
an important and fundamental minimum unit of length. Indeed, even  in the theory of the Dirac equation\cite{r15} every electron
apparently has the velocity of light $c$ brought out by the well known
Zitterbewegung or rapid oscillation effect and non-Hermitian (or complex valued)
position coordinate:
\begin{equation}
\vec x = (c^2 \vec p H^{-1} t ) + \frac{\imath}{2} c\hbar (\vec a -
c\vec p H^{-1})H^{-1},\label{e2}
\end{equation}
It is only on averaging over space-time intervals of the order of the Compton
wavelength and time that we recover physical electrons. This has been the basis
of the recent formulation of elementary particles as Kerr-Newman type
Black Holes\cite{r16,r17}
Though on the one hand the puzzling fact has been known that the Kerr-Newman metric
describes the field of the electron accurately, including the anomalous
$g=2$ factor\cite{r18}, on the other, it was recognized that an electron could not
be treated as a Kerr-Newman Black Hole as this would lead to a naked
singularity, that is a complex radius:
\begin{equation}
r_+ = \frac{GM}{c^2} + \imath b, b \equiv (\frac{G^2Q^2}{c^8} + a^2 -
\frac{G^2M^2}{c^4})^{1/2}\label{e3}
\end{equation}
Infact, as has been discussed in detail elsewhere (cf.refs.\cite{r12,r14,r16,r17}),
(\ref{e2}) corresponds to (\ref{e3}), and becomes meaningful once the above
averaging over these minimum space-time intervals is done - physics begins
outside these minimum intervals. It has also been shown that this leads to
a unified description of interactions - the quarks and leptons (including the
neutrino) are what may be called Quantum Mechanical Kerr-Newman Black Holes.
The quark description emerges at the Compton wavelength while outside we
recover the leptonic description\cite{r19,r20}.\\
It may be mentioned that such a minimum time interval, the chronon, has been
considered earlier in a different context by several authors\cite{r21,r22,r23}.\\
We now come to Nelson's derivation of the Schrodinger equation from Brownian
processes (cf.ref.\cite{r2}). In this case, the random change in the $x$
coordinate say, is given by,
$$| \Delta x| = \sqrt < \Delta x^2 > = \nu \sqrt \Delta t,$$
in an obvious notation, where the diffusion constant $\nu$ is given by
\begin{equation}
\nu = \frac{\hbar}{m} = lv\label{e4}
\end{equation}
$l$ being the mean free path or correlation length.\\
The adhoc identification of $\nu$ in (\ref{e4}) has been the troublesome
feature. This has been discussed elsewhere\cite{r24} (cf.also.ref.\cite{r16}),
but the point is that (\ref{e4}) gives us the Compton wavelength again.\\
The relativistic generalization of the above to the Klein-Gordon equation has been
even more troublesome\cite{r5}. In this case, there are further puzzling features apart from the
luminal velocity as in the Dirac equation, brought out by equation (\ref{e2}).
For Lorentz invariance, a discrete time is further required. Interestingly,
Snyder had shown that discrete space-time is compatible with Lorentz
transformations\cite{r25}. Here again, the Compton wavelength and time cut off
as discussed after (\ref{e2}) and (\ref{e3}) makes the whole picture transparent.\\
The stochastic derivation of the Dirac equation introduces a further
complication\cite{r6}. There is a spin reversal with the frequency $mc^2/\hbar$.
This again is readily explainable in the above context in terms of the Compton
time. Interestingly the resemblance of such a Weiner process to the Zitterbewegung
of the electron was noticed by Ichinose\cite{r26}.\\
Thus in all these cases once we recognize that the Compton wavelength and time
are minimum cut off intervals, the obscure or adhoc features become meaningful.\\
We would like to point out that the origin of the Compton wavelength is the
random walk equation (\ref{e1})! One could then argue that the Compton time
(or Chronon) automatically follows. This was shown by Hakim\cite{r27}.
We would like to point out the simple fact that a discrete space would automatically
imply discrete time. For, if $\Delta t$ could $\to 0$, then all velocities,
$lim_{\Delta t \to 0} |\frac{\Delta x}{\Delta t}|$ would $\to \infty \mbox{as}
|\Delta x |$ does not tend to $0!$. So there is a maximal velocity and this
in conjunction with symmetry considerations can be
taken to be the basis of special relativity.\\
Infact one could show that quantized
space-time is more fundamental than quantized energy and indeed would lead to the
latter\cite{r12}. To put it simply the frequency is given by $c/\lambda$, where $\lambda$
the wavelength is itself discrete and hence so also is the frequency. One could
then deduce Planck's law. This ofcourse, is the starting point of Quantum
Theory itself.\\
Equally interesting is the fact that a coordinate shift on Minkowski space-time
actually leads to the Dirac equation\cite{r28} on the one hand, and a
quantized picture of electromagnetism on the other\cite{r29}.\\
At this stage we make the following remark which throws further light
on the origin of the Compton wavelength (cf.ref.\cite{r14}). In the case of
the Dirac electron whose position is given by (\ref{e2}), the point
electron has the velocity of light and is subject to Zitterbewegung or rapid
oscillation within the Compton wavelength. The thermal wavelength for such a
motion is given by
$$\lambda = \sqrt \frac{\hbar^2}{mkT} \sim \mbox{Compton \quad wavelength}$$
by virtue of the fact that now $kT \sim mc^2$ itself.\\
Infact it has been shown in\cite{r16,r14}and\cite{r17}, that it is this circumstance that
leads to inertial mass, gravitation and electromagnetism (as for example brought out by the Kerr-Newman metric).
Other interactions also can be accommodated within this framework (cf.ref.\cite{r12}).
\section{Cosmological Considerations}
Given the minimum space-time intervals of the order of Compton wavelength
and time taking as usual the pion to be a typical elementary particle and
using the fact that given $N$ particles, there is a fluctuational creation
of particles of the $\sim \sqrt{N}$, a cosmological scheme has been
discussed in detail elsewhere, (cf.refs.\cite{r14}and \cite{r13} amongst
others for details). We merely report the fact that it deduces from theory
the age, radius and mass of the universe as also Hubble's law and other
hitherto inexplicable, purely adhoc relations like the mysterious relation
between the pion mass and the Hubble constant and the so called Dirac Large
Number relations including equation (\ref{e1}). The model also predicts an
ever expanding universe as indeed has since been confirmed\cite{r30}.
This cosmology shares certain features of the Prigogine, Dirac and Steady
State cosmologies.\\
For example, we have
$$\frac{dN}{dt} = \sqrt{N}/\tau$$,
where $\tau$ is the pion Compton time and
this leads to
\begin{equation}
T = \tau \sqrt{N}\label{e5}
\end{equation}
$T$ is the age of the universe. Other relations also follow as mentioned above.
The interesting point is, that (\ref{e5}) provides an arrow of time, whereas the
usual laws of physics, eg.electromagnetism are arrowless. As is now generally recognized an arrow
emerges from a thermodynamic (or irreversible) basis.\\
The point to be made here is that all this is very much in the spirit and letter
of the foregoing considerations. $N$, the number of particles in the
universe, typically pions, is the sole cosmological or large scale parameter,
while the microphysical parameter, the Compton wavelength (and time) also
follow from (\ref{e1}) - a circumstance that has been called stochastic
holism (cf.ref.\cite{r19}).
\section{Discussion}
We have attempted to realize Wheeler's "law without law" referred to in the
introduction. Wheeler (loc.cit.) speaks of a "regularity principle" which
added to a totally chaotic universe leads to law. For example, he considers
in the example of the travelling salesman, a "minimum distance" travelled, but
qualifies it by calling it a practical man's minimum (rather than a strict
mathematical minimum). In the above considerations, on the other hand, we
use the "average" distance (or time) in the sense of the "root mean square"
average.\\
So the "iron-clad" laws of the universe, have a stochastic underpinning.
Strictly speaking, these rigid laws emerge in the thermodynamic limit,
$N \to \propto$ (cf.ref.\cite{r31}). Ofcourse,
the laws are a good approximation because $N \sim 10^{80}.$ This should be
true of space and time reversal symmatrices also, in view of their discrete
structure, though the continuum is a good approximation.

\end{document}